 \documentclass[prl,twocolumn,showpacs,preprintnumbers,amsmath,amssymb]{revtex4}

\usepackage[dvips]{graphicx}
\usepackage{dcolumn}
\usepackage{bm}

\begin{document}


\title{  
           Addition Energies of Fullerenes and Nanotubes as
           Quantum Dots: \\
           The Role of Symmetry 
}

\author{ San-Huang Ke,$^{1,2}$ Harold U. Baranger,$^{2}$ and Weitao Yang$^{1}$}

\affiliation{
     $^{\rm 1}$Department of Chemistry, Duke University, Durham, NC 27708-0354 \\
     $^{\rm 2}$Department of Physics, Duke University, Durham, NC 27708-0305
}

\date{\today}

\begin{abstract}

Using density-functional theory calculations, we investigate the addition energy (AE) of quantum dots formed of fullerenes or closed single-wall carbon nanotubes. We focus on the connection between symmetry and oscillations in the AE spectrum.
In the highly symmetric fullerenes the oscillation period is large because of the large level degeneracy and Hund's rule. For long nanotubes, the AE oscillation is fourfold.  Adding defects destroys the spatial symmetry of the tubes, leaving only spin degeneracy; correspondingly, the fourfold behavior is destroyed, leaving an even/odd behavior which is quite robust.
We use our symmetry results to explain recent experiments. 
[Phys. Rev. Lett. {\bf 91}, 116803 (2003).]

\end{abstract}

\maketitle


Quantum dots (QDs) have been one of the most active research areas in condensed matter physics because of their fundamental interest as controllable electronic systems and their importance to potential applications such as quantum computing.  Two-dimensional (2D) QDs formed in semiconductor heterostructures have been studied intensively for a decade \cite{2d}.  Recently, new systems, such as nanotubes (1D) and nanoparticles or fullerenes (0D), became available experimentally and have been attracting more and more attention \cite{01d}. An important difference between the molecule-based and heterostructure-based QDs is that the molecules have a known atomic structure with definite symmetry. Here we show that this symmetry plays an important role in transport through the dots.

Our tool for probing symmetry in molecular QDs is the Coulomb blockade of transport in the tunneling regime \cite{2d,01d}: the electrostatic energy of an additional electron on a dot blocks the flow of current through the dot.  As a result, the conductance through the dot depends strongly on gate voltage, forming a series of sharp peaks. The spacing between these Coulomb blockade peaks is proportional to the addition energy (AE) -- the second difference of the ground state energy with respect to electron number. The shape of 2D semiconductor QDs is generally irregular.  Their levels and wavefunctions can either be treated statistically by random matrix theory \cite{Alhassid00} or be calculated by density functional theory (DFT) with a model irregular effective potential \cite{dft2d}.  In this way, most of the experiments on Coulomb blockade peak spacings in these dots can be understood statistically \cite{baranger1,baranger2}.

In contrast, the regular atomic structure of the new 0D and 1D QDs has high spatial symmetry, implying orbital degeneracies and so quite different behavior.  For example, for a long closed single-wall carbon nanotube (SWCNT), each orbital should be fourfold degenerate because of the combination of $K$-$K'$ valley degeneracy and spin. Therefore, a fourfold shell filling is expected, in contrast to the twofold even/odd effects observed in the AE of semiconductor quantum dots.  Recently, some non-tunneling experiments indeed observed four-electron shells \cite{fourfold1,fourfold2}. A later tunneling experiment found only twofold spin shells \cite{twofold}; in fact, a remarkably strong even/odd oscillation in the AE spectra was observed. Other experiments, however, did not observe even this twofold even/odd oscillation \cite{no_even/odd}. 

So far it is not clear how to reconcile these different experiments.  Indeed, for the new types of QDs many basic questions need to be addressed: What are the differences between the 0D and 1D systems? What is the effect of native defects?  What is the environmental influence under real experimental conditions?  To answer these questions unambiguously and to explain the different experiments, calculations free from empirical parameters are needed.

Here we report the first first-principles investigation of the AE spectra of fullerenes and closed SWCNTs.  Our main findings are: (1) For fullerenes there is no even/odd oscillation in the AE spectra because of the highly degenerate energy levels and Hund's rule -- the shell filling is multifold.  (2) For closed SWCNTs, the observability of periodic behavior depends on the energy window. Around the neutrality point, there is a strong even/odd oscillation, and it is fourfold for long clean tubes. Far from the neutrality point, the regular oscillation tends to vanish.  (3) We investigate the effect of both native defects and disturbance by an electrostatic potential. Defects destroy the fourfold oscillation, reducing it to twofold, while the effect of a potential disturbance is only minor.  The overall even/odd behavior around the neutrality point is surprisingly robust.  

All our findings can be explained by the electronic structure symmetry and its robustness: By causing degenerate levels, the underlying symmetry provides the necessary condition for an oscillation in AE. Whether this symmetry is observed depends on its robustness, both upon addition of electrons and upon other pertubations such as native defects. In this way we show that symmetry provides a reasonable explanation of the various experimental results.

Our calculations are carried out using the SIESTA package, a first-principles code of full DFT with a flexible and finite-range numerical LCAO basis set \cite{siesta}.  In our calculation, the PBE version of the generalized gradient approximation \cite{pbe} is adopted for the electron exchange and correlation, and optimized Troullier-Martins pseudopotentials \cite{tmpp} are used for the atomic cores.  

Two main technical difficulties have to be overcome. First, the self-consistency convergence is difficult because of the degenerate or nearly degenerate levels (even when using advanced charge mixing \cite{pulay}). We overcome this by introducing an electronic temperature: the result of an initial calculation at high temperature is used as input to a lower temperature calculation (and the process may be repeated)\cite{add}. In this way self-consistency can be achieved on the one hand, and the results are kept reliable on the other.  Second, for an AE calculation using periodic boundary conditions, we have to treat the added charge properly. A uniform charge background is introduced to compensate the net charge, a Madelung correction term is added to the total energy, and a simple cubic supercell is adopted so that the quadrupole contribution vanishes by symmetry.  This treatment turns out to be successful for AE calculations \cite{dmol}.  In our calculation different ground-state spin configurations are fully considered for each charge state.

An important parameter for interpreting the AE spectra is the ratio between the exchange shift and mean level spacing, $J/\Delta$.  There is a recent model calculation of this ratio for SWCNTs \cite{ratio}, but first-principles results are not, to our knowledge, available.  Here we propose a method to calculate this parameter motivated by a simplified two-level system:

\begin{figure}[h]
\includegraphics[angle=0,width=3.5cm]{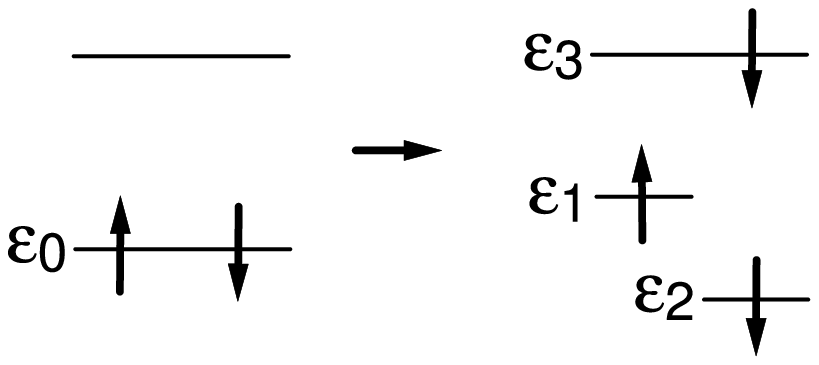}
\end{figure}

\noindent 
Consider a closed-shell (charged) system and add to it an electron. Because of the exchange interaction, the top spin-degenerate level ($\epsilon_0$) splits into two nondegenerate levels ($\epsilon_1$ and $\epsilon_2$).  The splitting is determined by the exchange interaction ($J$) and Coulomb interaction ($C$) between electrons 1 or 2 and 3: $\epsilon_1 = \epsilon_0 + C_{1,3}$, $\epsilon_2 = \epsilon_0 + C_{2,3} - J_{2,3}$.  Because the Coulomb energies $C_{1,3}$ and $C_{2,3}$ are approximately equal, the splitting is approximately the exchange shift $J$. The mean level spacing $\Delta$ is determined by averaging the level spacings within the energy range of interest.  Obviously, when $J/\Delta$ is larger than $\sim$ 0.5, the ordering of the spin shells will be destroyed by the exchange interaction.



\begin{figure}
\includegraphics[angle=0,width=8.0cm]{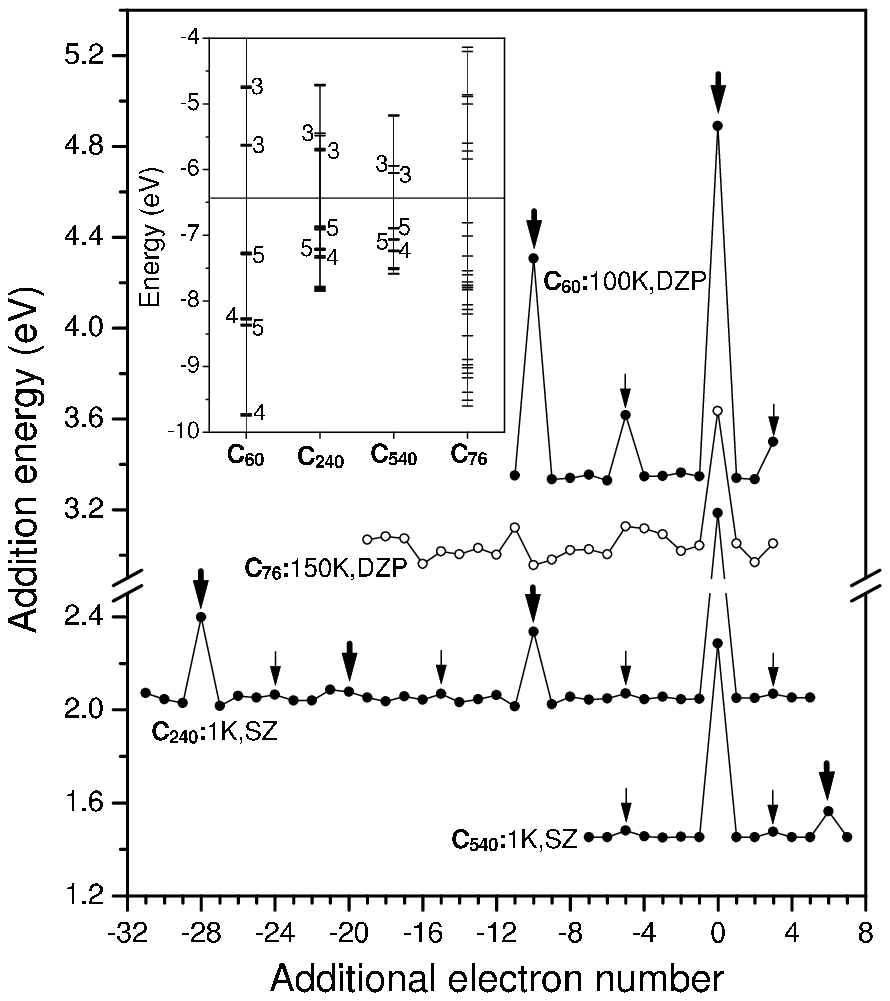}
\caption{Addition energy (AE) spectra of the fullerenes C$_{60}$, C$_{76}$, C$_{240}$, and C$_{540}$. 
Electronic temperatures and basis sets adopted are indicated.
Inset: the levels near the neutrality point, showing the large degeneracy factors.
Note the lack of any even/odd oscillation in the AE; 
the large values occur upon completion of a shell.} 
\end{figure}

{\it Fullerenes} -- The balls we have calculated include C$_{60}$, C$_{76}$, C$_{240}$, and C$_{540}$; the results are summarized in Fig. 1.  Because of similar spatial symmetry ($I_h$), the level structures of C$_{60}$, C$_{240}$, and C$_{540}$ near the neutrality point are similar: 5-fold degeneracy for the highest occupied molecular orbital and 3-fold for the lowest unoccupied molecular orbital (Fig. 1 inset).  As a result, the AE spectra of these three balls is similar.  Hund's rule is fully obeyed as electrons are added or removed: high spin configurations (up to $S=5/2$) are always preferred because of the large level degeneracy in each shell (up to 10). There is a weak peak (thin arrow in Fig. 1) at the middle of each shell where electrons with opposite spin begin to fill -- a result of the slightly larger on-site Coulomb energy -- and a strong peak (solid arrow in Fig. 1) between two adjacent shells because of the large inter-shell level spacing.  Consequently, there is no regular even/odd oscillation in the AE spectra.

Among the five fullerene balls, C$_{76}$ has a lower spatial symmetry ($D_2$), leading to a more distributed level structure (see inset to Fig. 1).  However, because of the clustering of many nearly degenerate levels (around $-7.8$ eV, for instance), there is no regular even/odd oscillation.  This clustering leads to even higher spin configurations (up to $S=7/2$) and a very large value of $J/\Delta$ ($\sim$1.2).  
Under these conditions, an unusual ``anti-even/odd" oscillation in the AE spectra -- peaks for odd numbers of electrons and valleys for even -- sometimes occurs, for instance around electron number $-12$.

We have carefully studied how the results in Fig. 1 are influenced by three computational factors: (1) electronic temperature, (2) size and type of basis set, and (3) charge-induced structure change. We find that the effects of these factors are small \cite{add}.

Experimentally, it is now possible to measure tunneling transport through a single fullerene molecule: recently Coulomb blockade effects were observed in a single C$_{60}$ ``molecular transistor'' \cite{c60_trans}. Although no experimental AE spectra of fullerenes are available to date, our theoretical results shed light on possible future experiments. 


\begin{figure}
\includegraphics[angle=0,width=8.0cm]{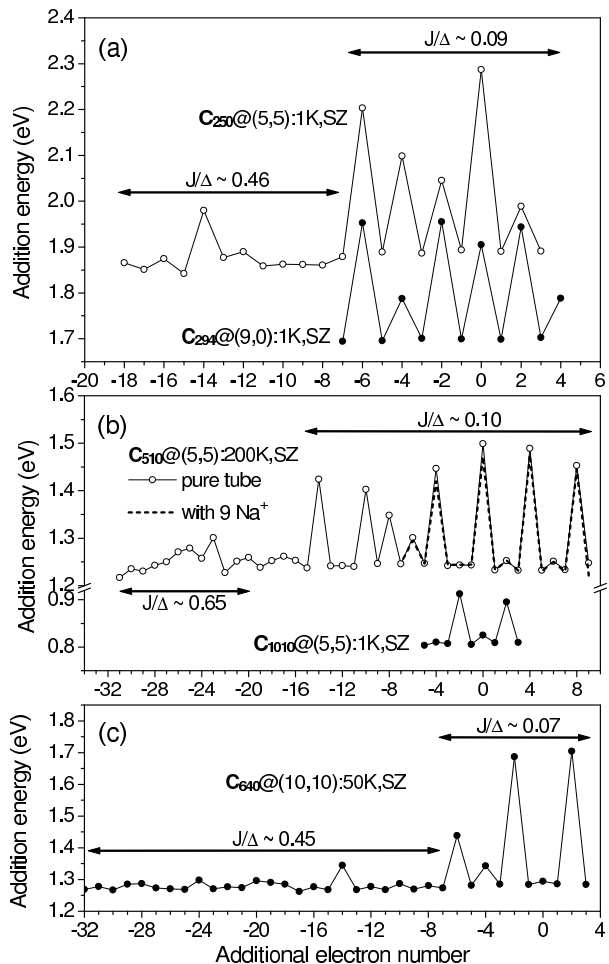}
\caption{AE spectra of the closed SWCNTs (a) C$_{250}$ and C$_{294}$, 
(b) C$_{510}$ and C$_{1010}$, and (c) C$_{640}$. 
Note the large even/odd oscillation near the neutrality point in 
the small tubes and the fourfold oscillation for the longer tubes.
Values of $J/\Delta$ (ratio of exchange shift to mean-level spacing) are calculated by averaging over the levels indicated by the arrows.
}
\end{figure}

{\it Perfect carbon nanotubes} -- We have calculated AE spectra for the following closed SWCNTs: the (5,5) tubes C$_{250}$, C$_{510}$, and C$_{1010}$ (capped by C$_{60}$); the (9,0) tube C$_{294}$ (capped by C$_{78}$); and the (10,10) tube C$_{640}$ (capped by C$_{240}$). Fig. 2 shows the main features: (1) near the neutrality point there are strong even/odd oscillations, (2) for short tubes (C$_{250}$ and C$_{294}$) the oscillation is twofold while for longer tubes (C$_{510}$ and C$_{1010}$) it is fourfold, and (3) far from the neutrality point the even/odd oscillation tends to vanish.

Further analysis reveals that the strong even/odd oscillation near the neutrality point is due to the large mean level spacing $\Delta$ -- much larger than the exchange-induced level shift $J$ (see Fig. 2 for values of $J/\Delta$).  This large $\Delta$ originates from the low density of states in infinitely-long metallic nanotubes (see Fig. 3).  For very long tubes, the $K$ and $K'$ valleys are degenerate -- a result of the equivalence of the two atoms in the primitive cell of graphene -- so each level should be fourfold degenerate (including spin).  For the short tubes (C$_{250}$ and C$_{294}$) this fourfold degeneracy is destroyed  by  the symmetry breaking caused by the caps [Fig. 2 (a)].  As the tube gets longer (from C$_{250}$ to C$_{510}$ and finally to C$_{1010}$), the effect of the caps becomes weaker.  Consequently, we see nearly fourfold oscillations in Fig. 2 (b) for C$_{510}$ and C$_{1010}$.

Our finding is consistent with a recent experiment showing four-electron periodicity in the conductance as a function of gate voltage \cite{fourfold1}.  However, the nanotube in that experiment was an ``open" device in that the contacts were transparent.  For closed nanotube dots, only twofold spin-shell filling has been found \cite{twofold}, which has not been understood so far.  We show below that the effect of native defects can explain this observation.

\begin{figure}
\includegraphics[angle=0,width=7.0cm]{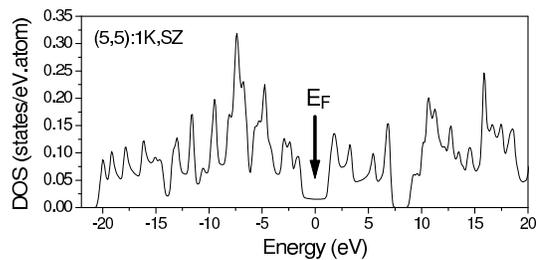}
\caption{Density of states of an infinite (5,5) nanotube. 
The low value at the neutrality point implies a large mean-level spacing in the closed tubes.}
\end{figure}

Even the longest tube considered here (C$_{1010}$, 12.4 nm long) is still much shorter than those in experiments (tens to hundreds of nanometers) \cite{fourfold1,fourfold2,twofold}.  As the tube becomes longer, $\Delta$ becomes smaller, and one should question whether the strong even/odd oscillation near the neutrality energy survives.  Because the real tubes are impossible to handle computationally, we calculate $J/\Delta$ near the neutrality energy for the C$_{250}$, C$_{510}$, and C$_{1010}$ tubes, and extrapolate the trend.  The result is $J/\Delta =$ 0.09, 0.10, and 0.10, respectively.  Because the ratio does not vary with length, a strong even/odd effect near the neutrality energy should still be present in the very long tubes in real experiments.

Far from the neutrality energy, Fig. 3 shows that the density of states increases rapidly, causing a rapid increase in $J/\Delta$ to $\sim 0.5$ (see Fig. 2).  Consequently, the regular even/odd behavior tends to vanish, similar to the case of the fullerenes.  Of course, for the nanotubes in current experiments this regime corresponds to adding a very large amount of charge.

We have carefully checked the above results for the influence of computational factors, as for the fullerenes; for details see Ref. \onlinecite{add}.

\begin{figure}
\includegraphics[angle=0,width=8.0cm]{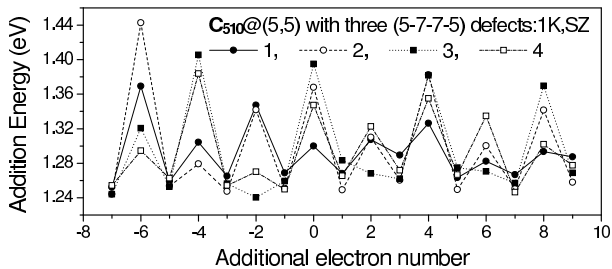}
\caption{AE spectra of the C$_{510}$ tube with three (5-7-7-5) defects
separated widely and orientated randomly.
1, 2, 3, and 4 denote four different defect configurations. 
The presence of defects destroys the fourfold periodicity, 
but a strong even/odd effect remains.}
\end{figure}


{\it Imperfect carbon nanotubes} -- Up to now, we have discussed ideal closed tubes.  In real experiments, however, nanotubes always have some defects.  In addition, in most transport experiments a nanotube QD is in an environment of external charges (trapped in the insulating substrate, for instance);  therefore, there may be effects due to an external electrostatic potential \cite{point_charge}.

To investigate the defect effect, we consider the native (5-7-7-5) defect \cite{5775_1, 5775_2} which is formed under tension by rotating one C-C bond by 90$^{\circ}$.  Experimentally, the concentration of this (5-7-7-5) defect varies depending on the preparation process; as a rough estimate, 1/(hundreds of atoms) seems reasonable \cite{liu}. We introduced three (5-7-7-5) defects into the C$_{510}$ tube at widely separated locations and with random orientations. All the structures were relaxed to equilibrium.  

Four calculated AE spectra are shown in Fig. 4.  Different defect configurations have, of course, different effects, resulting in wide distributions of both the baseline energy (charging energy) and the amplitude of even/odd oscillation.  Two interesting facts are (1) the even/odd oscillation is quite robust -- it survives for most defect configurations -- and (2) the defects destroy the fourfold shell filling in the longer tubes and reduces it to twofold.  The second result explains why only twofold shell filling was observed in the tunneling transport experiment \cite{twofold}.

To see the influence of an external electrostatic potential, we attached 9 Na atoms randomly onto the surface of one side of the C$_{510}$ tube.  Because of the large difference in electron affinity between C and Na (the $2p$ electrons are included as valence electrons), almost one full ($3s$) electron is transferred from each Na atom to the tube. The Na atoms, then, will act as point charges near the tube. In this way, we created an electrostatic potential disturbance for the tube.  The calculated AE spectra are shown in Fig. 2 (b) (dashed line).  As can be seen, the effect is almost negligible.  Because the concentration of external charges is experimentally unknown, this small effect only shows that it is much weaker than the defect effect for comparable concentrations.

 

This work was supported in part by the NSF (DMR-0103003).


\begin{references}
\bibitem{dmol} As a test we calculate the C$_{250}$@(5,5) tube using SIESTA and 
               DMol3 package (with non-periodic boundary condition). 
               The calculated AE spectra are in perfect agreement except for 
               a small constant shift \cite{add}.
\end{references}
\end{document}